\documentclass[12pt]{article}
\usepackage{amsmath}
\usepackage{cite}
\usepackage{relsize}
\usepackage{amssymb}
\usepackage{graphics}
\usepackage{epsfig}
\usepackage{epstopdf}
\usepackage{verbatim}
\usepackage{color}
\usepackage{subcaption}
\usepackage{multirow}
\usepackage{textcomp}
\usepackage{eqnarray}
\usepackage{float}
\usepackage{mathtools}
\usepackage{pifont}
\newcommand{\cmark}{\ding{51}}%
\newcommand{\xmark}{\ding{55}}%
\usepackage[toc,page]{appendix}
\usepackage{enumitem}
\usepackage{ragged2e}
\usepackage{dirtytalk}

\begin{document}

\title{\textbf{Scotogenesis in Hybrid Textures of Neutrino Mass Matrix and Neutrinoless Double Beta Decay}}

\author {Ankush\thanks{ ankush.bbau@gmail.com}, Monal Kashav\thanks{ monalkashav@gmail.com}, Surender Verma\thanks{ s\_7verma@yahoo.co.in}  and B. C. Chauhan\thanks{ bcawake@hpcu.ac.in}}

\date{\textit{Department of Physics and Astronomical Science,\\Central University of Himachal Pradesh, Dharamshala 176215, INDIA.}}

\maketitle
\begin{abstract}
\noindent We study the connection between dark matter (DM) and neutrinoless double beta ($0\nu\beta\beta$) decay in a scotogenic model with hybrid texture in the neutrino mass matrix. Characteristically, the framework allows to write all the non-zero elements of the mass matrix proportional to effective Majorana mass $\left|M_{ee}\right|$. The overall scale of the neutrino mass is found to be governed by non-zero $\left|M_{ee}\right|$. We have obtained interesting correlations of relic density of DM($\Omega h^2$) with DM mass $M_1$ and $\left|M_{ee}\right|$. Using experimental value of DM relic density($\Omega h^2$), the DM mass $M_1$, is found to be $\mathcal{O}$(1TeV) which is within reach of collider experiments. Specifically, for all five hybrid textures, the range of upper bound on DM mass $M_1$ is found to be ($2.27$-$5.31$)TeV. Another interesting feature of the model is the existence of lower bound on $|M_{ee}|$ for all allowed hybrid textures except texture $T_5$ which can be probed in current and future $0\nu\beta\beta$ decay experiments. With high sensitivities, these experiments shall establish the theoretical status of the proposed model. For example, the non-observation of $0\nu\beta\beta$ decay down to the sensitivity $\mathcal{O}(0.03)$eV will refute $T_3$ hybrid texture.      
\end{abstract}
\section{Introduction}
After the experimental observation of neutrino oscillations\cite{Kam11,sno1}, the immediate quest is to understand the origin of neutrino mass eventually responsible for observed mixing pattern in the leptonic sector. Despite astounding success standard model (SM) of particle physics is inadequate to explain non-zero neutrino mass. Various beyond SM scenarios have been proposed providing a coherent and unified theoretical structure to understand the underlying dynamics of neutrino mass generation. Vast majority of these models are based on tree-level seesaw frameworks\cite{Minkowski:1977sc,Mohapatra:1979ia,TY,Glashow:1979nm,Gell11}. Alternatively, radiative neutrino mass models are very appealing as they can accommodate solar and atmospheric neutrino mass scales by suppression emanating from the loop factors and relatively lighter mass of the mediator particle which may be observed at current collider experiments. 

\noindent On the other hand, there is another apparently independent unsettled problem about the nature of the dark matter (DM). The radiative models forged with discrete symmetries such as $Z_2$ provide a natural way to stabilize the viable DM candidate(s). In this way, these extensions, beyond tree-level dynamics popularly known as ``\textit{scotogenic models}", offer promising explanation to both non-zero neutrino mass and nature of DM. In these models the SM field content is enhanced by the addition of three singlet fermions and a scalar doublet \cite{Ma1}.

\noindent Another fundamental problem in neutrino physics pertains to discerning Dirac or Majorana nature of neutrinos. Such identification is very important for neutrino mass model building. In fact, these models blooms from the very basic assumption about the nature of neutrino being Dirac or Majorana. One of the characteristic feature of Majorana neutrino is the possibility for existence of lepton number violating (LNV) processes such as neutrinoless double beta ($0\nu\beta\beta$) decay. Although this process is still elusive but, once observed, will confirm Majorana nature of neutrino.

\noindent  Radiative generation of neutrino mass can be done at one-loop level \cite{Ma2,Zee:1980ai,Ma:2001mr,Kubo:2006yx,Hambye:2006zn,Farzan:2009ji}, two-loop level \cite{Ma:2007yx,Ma:2007gq,Kajiyama:2013zla,Aoki:2013gzs,Kajiyama:2013rla}, three and higher-loop level \cite{Krauss:2002px,Aoki:2008av,Gustafsson:2012vj,Ahriche:2014cda,Ahriche:2014oda,Nomura:2016seu}. The basic scotogenic model proposed by E. Ma is the most discussed radiative  models at one-loop level, as this model simultaneously account for the neutrino mass generation and dark matter \cite{Ma1}. In these models, additional suppression is guaranteed by the loop integrals, and this suppression is higher for higher loops. Therefore, with higher-loops, we can expect charged scalar to be lighter enough to be tested at LHC. Scotogenic model assuming texture zeros in the neutrino mass matrix has been discussed in Refs. \cite{Kitabayashi:2018bye,Kitabayashi:2017sjz}.

\noindent In this work, we attempt to integrate the explanation to aforementioned problems into a single framework assuming neutrino to be Majorana particle. Within scotogenic model, we work in a more constrained scenario wherein hybrid textures are assumed in the neutrino mass matrix. Hybrid texture means one-zero and one-equality amongst  the elements of neutrino mass matrix.  One of the important property of hybrid textures  considered in this work is that all the elements except \say{0-element} can be written, in a way, proportional to non-zero $\left|M_{ee}\right|$ element of the neutrino mass matrix.

\noindent The paper is organized as follows. In Section 2 we outline the basic structure of scotogenic model. In Section 3, we discuss the co-annihilation of dark matter and calculation of relic density of DM in the current setup. In Section 4, we discuss the connection between DM and $0\nu\beta\beta$ decay assuming hybrid textures in the neutrino mass matrix. The details of the numerical analysis and interpretation of the phenomenological results are discussed in Section 5. Finally, the conclusions are elucidated in Section 6.
\section{The model}
\par The scotogenic model  is an extension of the standard model (SM) wherein three  Majorana $SU(2)_{L}$ singlet  fermions $N_{k}  (k=1,2,3)$ and  an   $SU(2)_{L}$ scalar doublet $(\eta^{+},\eta^{0})$  are added to the standard field content \cite{Ma1}. All these beyond standard model particles are odd under exact $Z_{2}$ symmetry. The  particle content of the scotogenic model under $SU(2)_{L}\times U(1)_{Y} \times Z_{2}$ is given by \\
\begin{equation}
\begin{split}
    L_{\alpha}&=(\nu_{\alpha},l_{\alpha}):(2,-1/2,+), l^{C}_{\alpha}:(1,1,+),\\
\phi&=(\phi^{+},\phi^{0}):(2,-1/2,+),\\
\eta&=(\eta^{+},\eta^{0}):(2,1/2,-),N_{k}:(1,0,-),
\end{split}
\end{equation}
where $\alpha=e,\mu,\tau$,  $(\nu_{\alpha},l_{\alpha})$ and $(\phi^{+},\phi^{0})$ are left-handed lepton doublets and Higgs doublet, respectively.\\
The Lagrangian of the scotogenic model containing relevant Yukawa and mass terms is given by
\begin{equation}
    \mathcal{L} \supset y_{\alpha k}(\bar{\nu}_{\alpha L}\eta^{0}-\bar{l}_{\alpha L} \eta^{+})N_{k} +\frac{1}{2}M_{k}\bar{N}_{k}N^{C}_{k} +H.c.,
\end{equation}
and the relevant interaction terms in the scalar potential are given by
\begin{equation}
    V\supset \frac{1}{2}\lambda (\phi^{\dagger}\eta)^{2} +H.c.,
\end{equation}
where $\lambda$ is the quartic coupling.
Due to exact $Z_{2}$ symmetry neutrino mass at tree level is forbidden and they will acquire mass via one-loop level.\\
After integrating heavy fields at one-loop level the general element of neutrino mass matrix \say{$M_{\alpha \beta}$} is given by 
\begin{equation}
\label{I}
    M_{\alpha \beta} = \sum _{k=1}^{3} y_{\alpha k}y_{\beta k}\Lambda_{k},
\end{equation}
where
\begin{equation}
    \Lambda_{k}=\frac{\lambda v^{2}}{16 \pi^{2}}\frac{M_{k}}{m_{0}^{2}-M_{k}^{2}}\left(1-\frac{M_{k}^{2}}{m_{0}^{2}-M_{k}^{0}}\ln\frac{m_{0}^{2}}{M_{k}^{2}}\right),
\end{equation}
\begin{equation}
    m_{0}^{2}=\frac{1}{2}(m_{R}^{2}+m_{I}^{2}),
\end{equation}
and $v$, is the vacuum expectation value ($vev$) of the Higgs field, $m_{R}$ and $m_{I}$ are the masses of $\sqrt{2} Re[\eta^{0}]$ and $\sqrt{2} Im [\eta^{0}]$, respectively and $M_{k} (k=1,2,3)$ are right-handed neutrino masses.
\noindent This model, also, accounts for the lepton flavor violating (LFV) processes such as $\mu\to e\gamma$ induced at one loop level.
The branching ratio for process $\mu\to e\gamma$ is given by \cite{Kubo:2006yx,Ma:2001mr}
\begin{equation}
    Br(\mu\to e\gamma)=\frac{3 \alpha_{em}}{64 \pi(G_{F}m_{0}^{2})^{2}}\left|\sum_{k=1}^{3}y_{\mu k}y_{e k}^{*}F\left(\frac{M_{k}}{m_{0}}\right)\right|^{2},
\end{equation}
where $\alpha_{em}$ is the fine structure constant for electromagnetic coupling,  $G_{F}$ is  Fermi coupling constant and $F(r)$ is given by
\begin{equation}
F(r)=\frac{1-6r^{2}+3r^{4}+2r^{6}-6r^{4}  \ln r^{2}}{6(1-r^{2})^{4}} , \hspace{0.2cm} r\equiv \frac{M_{k}}{m_{0}}.   
\end{equation}
\section{Co-annihilation of the cold dark matter and relic density}
 The characteristic feature of scotogenic model is that it provide a framework for simultaneous explanation of neutrino mass and DM. $N_{k}$ is odd under $Z_{2}$ symmetry. The lightest of  $N_{k}$ is, thus, the suitable dark matter candidate in the model. The co-annihilation of lightest $Z_{2}$ odd particle  ($N_{1}$), through Yukawa couplings account for the cold dark matter candidate abundance and, also, for branching ratios of flavor violating $\mu \to e \gamma$ process consistent with scotogenic model \cite{Griest}. Here we assume that the mass of dark matter candidate $N_{1}$ is almost degenerate with mass of next singlet fermion $N_{2}$. The overall right-handed neutrino mass spectrum here is $M_{1}\leq M_{2} < M_{3}$ \cite{Griest}.  The co-annihilation cross-section times the relative velocity of annihilating particles $v_{r}$ is given by \cite{Suematsu:2009ww}
\begin{equation}
\label{a1}
    \sigma_{ij}|v_{r}|=a_{ij}+b_{ij}v_{r}^{2},
\end{equation}
where

\begin{align}
\left.
 \begin{array}{lll}
     a_{ij}=\frac{1}{8\pi} \frac{M_{1}^2}{(M_{1}^2+m_{0}^{2})^{2}}\sum_{\alpha, \beta}(y_{\alpha i}y_{\beta j}-y_{\alpha j}y_{\beta i})^{2},\\
    b_{ij}=\frac{m_{0}^{4}-3m_{0}^{2}M_{1}^{2}-M_{1}^{4}}{3(M_{1}^{2}+m_{0}^{2})^{2}} a_{ij}+\frac{1}{12 \pi}\frac{M_{1}^{2}(M_{1}^{4}+m_{0}^{4})}{(M_{1}^{2}+m_{0}^{2})^{4}}\sum_{\alpha,\beta} y_{\alpha i}y_{\alpha j}y_{\beta i}y_{\beta j},
 \end{array}
 \right\}
 \end{align}
and $\sigma_{ij} (i,j=1,2)$ is the annihilation cross-section for the process $N_{i} N_{j}\to x x^{'}$, $d M = (M_{2}-M_{1})/M_{1}$ is the mass splitting ratio for the almost degenerate singlet fermions, $x=M_{1}/T$ i.e. ratio of lightest singlet fermion $N_{1}$ to the temperature T.
If $g_{1}$, $g_{2}$ are the number of degrees of  freedom of singlet fermions $N_{1}$ and $N_{2}$, respectively, the effective cross section is given by
\begin{equation}
\label{a2}
    \sigma_{eff}=\frac{g_{1}^{2}}{g_{eff}^{2}}\sigma_{11}+\frac{2g_{1}g_{2}}{g_{eff}^{2}}\sigma_{12}(1+dM)^{3/2}\exp(-xdM)+\frac{g_{2}^{2}}{g_{eff}^{2}}\sigma_{22}(1+dM )^{3}\exp(-2 xdM),
\end{equation}
\begin{equation}
\label{a3}
    g_{eff}=g_{1}+g_{2}(1+dM)^{3/2} \exp(-xdM),
\end{equation}
with $dM \simeq 0$ ($N_{1}$ is almost degenerate with $N_{2}$) and using Eqns.(\ref{a1}) and (\ref{a3}) in Eqn.(\ref{a2}) we get
\begin{equation}
\sigma_{eff}|v_{r}|=\left(\frac{\sigma_{11}}{4}+\frac{\sigma_{12}}{2}+\frac{\sigma_{22}}{4}\right)|v_{r}|,
\end{equation}
  \begin{equation*}
      =a_{eff}+b_{eff}v_{r}^{2},  
  \end{equation*}
where 
\begin{align}
    \left.
    \begin{array}{cc}
      a_{eff}=\frac{a_{11}}{4}+\frac{a_{12}}{2}+\frac{a_{22}}{4},\\
          b_{eff}=\frac{b_{11}}{4}+\frac{b_{12}}{2}+\frac{b_{22}}{4}.
\end{array}
\right\}
\end{align}
The thermal average cross section is
\begin{equation}
\label{lebel1}
    <\sigma_{eff}|v_{r}|> = a_{eff}+6 b_{eff}/x,
\end{equation}
 which has linear dependence on temperature as $x=M_{1}/T$ and relic abundance of cold dark matter is given by
\begin{equation}
    \Omega h^{2}=\frac{1.07\times 10^{9} \text{GeV}^{-1}}{J g_{*}^{1/2} m_{Pl}},
\end{equation}
where $m_{Pl}=1.22\times10^{19}$GeV , $g_{*}=106.75$
and
\begin{equation}
    J(x_{f})=\int_{x_{f}}^ \infty \frac{<\sigma_{eff}|v_{r}|>}{x^{2}} dx,
\end{equation}
where $x_{f}=M_{1}/T_{f}\approx 25$, $T_{f}$ is the freeze-out temperature\cite{Kolb:1990vq}.
 \\ 
Using Eqn.(\ref{lebel1}), the relic density become
\begin{equation}
\label{omega}
    \Omega h^{2} =\frac{1.07\times 10^{9} \text{GeV}^{-1} x_{f}}{(a_{eff}+3 b_{eff}/x_{f}) g_{*}^{1/2} m_{Pl}},
\end{equation}
where
\begin{equation}
 x_{f}=\ln \frac{0.038 g_{eff} m_{Pl} M_{1}<\sigma_{eff}|v_{r}|> }{g_{*}^{1/2}x_{f}^{1/2} }.   
\end{equation}
\section{Connecting dark matter and $0\nu\beta\beta$ decay in the framework of hybrid textures}
After setting out the basic framework of the scotogenic model and modalities to calculate relic density of DM in earlier sections, here we analyse neutrino mass matrix
\begin{equation}
M_{\nu}= U diag (m_{1},m_{2}e^{i\alpha_2},m_{3}e^{i\alpha_3}) U^{T}\equiv\begin{pmatrix}
M_{ee} & M_{e\mu} & M_{e\tau}\\
M_{e\mu} & M_{\mu\mu} & M_{\mu\tau} \\
M_{e\tau} & M_{\mu\tau} &  M_{\tau\tau} 
\end{pmatrix},
\end{equation}
 \\
\noindent  focusing on the, yet elusive, neutrinoless double beta ($0\nu\beta\beta$) decay. Here $U$ is the Pontecorvo–Maki–Nakagawa–Sakata (PMNS) mixing matrix, ($m_1,m_2,m_3$) are three neutrino mass eigenvalues and ($\alpha_2,\alpha_3$) are Majorana phases. In general, $U$ is parameterized in term of three mixing angles ($\theta_{12},\theta_{23}, \theta_{13}$) and Dirac-type $CP$ violating phase $\delta$. In fact, the most important characteristic of Majorana nature of neutrino is the possibility of lepton number violating $0\nu\beta\beta$ decay. Beside the phase space factors, the amplitude of this decay is proportional to (1,1) element of neutrino mass matrix, ($\left|M_{ee}\right|$). The elements of the mass matrix are, in general, functions of three mixing angles, three mass eigenvalues and three $CP$ violating phases. Alternatively, we can rearrange, the elements of $\mu-\tau$ sector of the mass matrix to write them as  {\cite{Kitabayashi:2015tka, Kitabayashi:2015jdj}}
\begin{equation}
\label{lebel2}
M_{\mu\mu,\tau\tau, \mu\tau}=f(M_{ee},M_{e\mu},M_{e\tau},\theta_{12},\theta_{23},\theta_{13},\delta),
 \end{equation}
 for example,\\
 \begin{align*} 
 M_{\mu\mu} &= (A+B\cos{2\theta_{23}}) M_{ee}+((C+D)s_{23}+2s_{23}^{2}E)M_{e\mu}+((C+D)c_{23}+2c_{23}^{2}F)M_{e\tau},\\
 M_{\mu\tau} &= (A-B\cos{2\theta_{23}}) M_{ee}+((D-C)s_{23}-2s_{23}^{2}E)M_{e\mu}+((D-C)c_{23}-2c_{23}^{2}F)M_{e\tau},\\
  M_{e\tau} &= -B\sin{2\theta_{23}}M_{ee}-(Cs_{23}\tan{2\theta_{23}}+E)M_{e\mu}-(Cc_{23}\tan{2\theta_{23}}-F)M_{e\tau}.
\end{align*}
where
\begin{align*}
c_{ij}&\equiv\cos\theta_{ij},\\
s_{ij}&\equiv\sin\theta_{ij},\\
A &=  \frac{1}{2}(1+e^{-2 i \delta}),\\
B &=\frac{1}{2}(1-e^{-2i \delta}),\\
C &=-\frac{1}{2}\cos{2\theta_{23}} (e^{-i \delta}\cot{2\theta_{13}}-e^{i\delta}s_{13}c_{13}^{-1}),\\
D &=  e^{-i \delta}\cot{2\theta_{23}}-\frac{1}{2}e^{i\delta}s_{13}c_{13}^{-1},\\
E &=   c_{23}(\cot{2\theta_{12}}\sec{\theta_{13}}-e^{-i\delta}\csc{2\theta_{23}}s_{13}c_{13}^{-1}),\\
F &=s_{23}(\cot{2\theta_{12}}\sec{\theta_{13}}-e^{-i\delta}\csc{2\theta_{23}}s_{13}c_{13}^{-1}),
\end{align*}
($i,j=1,2,3; i<j$) and neutrino mass eigenvalue as
 \begin{equation}
\begin{split}
m_{1,2,3}=f(M_{ee},M_{e\mu},M_{e\tau},\theta_{12},\theta_{23},\theta_{13},\delta,\alpha_{2},\alpha_{3}).
\end{split}
\end{equation}

\noindent The rearrangement done in Eqn.(\ref{lebel2}) is general, however, in order to explain the possible connection between dark matter and $0\nu\beta\beta$ decay amplitude in the current setup, we consider a constrained class of models wherein hybrid textures are assumed in the neutrino mass matrix in such a way that element $M_{\alpha\beta}$, $((\alpha,\beta)=e,\mu,\tau)$, is proportional to non-zero $|M_{ee}|$. Hybrid texture imply one \textit{zero} and one \textit{equality} amongst the elements of neutrino mass matrix \cite{Kaneko,Dev:2009he,Goswami:2008uv,Liu:2013oxa,Kalita:2015tda}. Using Eqn.(21), it can be seen that there are nine possibilities to realize hybrid textures in $M_\nu ($with $|M_{ee}|\ne0$) tabulated in Table \ref{T1}. Out of nine only six ($T_{1....6}\equiv T_1, T_2, T_3, T_4, T_5$ and $T_6$) are consistent with experimental data on neutrino mass and mixing \cite{Dev:2009he} \textit{viz.}

 \begin{table}[t]
  \centering
\begin{tabular}{ |c|c|l| }
 \hline
Zero & Equality & Allowed(\cmark)/Disallowed (\xmark) \\
 \hline
 $M_{\mu\mu}=0$ &   & \cmark $\coloneqq T_1$\\
\cline{1-1} \cline{3-3}
$M_{\tau\tau}=0$ & $M_{e\mu}=M_{e\tau}$ & \cmark$\coloneqq T_2$\\
\cline{1-1} \cline{3-3}
 $M_{\mu\tau}=0$ &   & \xmark\\
 \hline
  & $M_{\mu\mu}=M_{\tau\tau}$ & \xmark \\ 
\cline{2-3}
$M_{e\mu}=0$  & $M_{\mu\mu}=M_{\mu\tau}$ & \cmark $\coloneqq T_3$\\ 
\cline{2-3}
 & $M_{\tau\tau} =M_{\mu\tau}$& \cmark$\coloneqq T_4$ \\
 \hline
   & $M_{\mu\mu}=M_{\tau\tau}$ & \xmark \\
 \cline{2-3}
$M_{e\tau}=0$ & $M_{\mu\mu}=M_{\mu\tau}$ & \cmark$\coloneqq T_5$ \\
 \cline{2-3}
  & $M_{\tau\tau}=M_{\mu\tau}$
 & \cmark$\coloneqq T_6$\\
 \hline

 \end{tabular}
 \caption{Nine possible hybrid textures for which $M_{\alpha\beta}$, $((\alpha,\beta)=e,\mu,\tau)$, is proportional to non-zero $\left|M_{ee}\right|$. The textures listed here are allowed(\cmark) or disallowed(\xmark) based on whether they reproduce correct low energy phenomenology or not.}
 \label{T1}
 \end{table}

\begin{center}
\vspace{0.4cm}
$T_{1}:\begin{pmatrix}
M_{e e} & \Delta & \Delta \\
- & 0 & M_{\mu \tau} \\
- & - &  M_{\tau \tau} 
\end{pmatrix}$, \hspace{0.4cm}
$T_{2}:\begin{pmatrix}
M_{e e} & \Delta & \Delta \\
- & M_{\mu \mu} & M_{\mu \tau} \\
- & - & 0 
\end{pmatrix}$,\\
\vspace{0.4cm}
$T_{3}:\begin{pmatrix}
M_{ee} & 0 & M_{e \tau}\\
- & \Delta & \Delta \\
- & - &  M_{\tau \tau} 
\end{pmatrix}$, \hspace{0.4cm}
$T_{4}:\begin{pmatrix}
M_{e e} & 0 & M_{e \tau}\\
- & M_{\mu \mu} & \Delta \\
- & - & \Delta 
\end{pmatrix}$,\\
\vspace{0.4cm}
$T_{5}:\begin{pmatrix}
M_{e e} & M_{e \mu} & 0 \\
- & \Delta & \Delta \\
- & - &  M_{\tau \tau} 
\end{pmatrix}$, \hspace{0.4cm}
$T_{6}:\begin{pmatrix}
M_{e e} & M_{e\mu} & 0\\
- & M_{\mu \mu} & \Delta \\
- & - &\Delta 
\end{pmatrix}$,
\end{center}

\noindent where $\Delta$ represents the equal elements. For these six hybrid textures, the six elements of neutrino mass matrix can be written as
\begin{equation}
\label{meq}
M_{\alpha\beta}=f^X_{\alpha\beta}(\theta_{12},\theta_{23},\theta_{13},\delta) M_{ee},
\end{equation} 
where ($\alpha,\beta=e,\mu,\tau$) and $X=T_{1....6}$. It is interesting to note that the overall scale of neutrino mass is governed by non-zero $M_{ee}$ i.e.

\begin{center}
$T_{1}:\begin{pmatrix}
1 & f_{e \mu}^{T_1} & f_{e \mu}^{T_1} \\
- & 0 & f_{\mu\tau}^{T_1} \\
- & - &  f_{\tau\tau}^{T_1} 
\end{pmatrix}M_{ee}$, \hspace{0.4cm}
$T_{2}:\begin{pmatrix}
1 & f_{e \mu}^{T_2} & f_{e \mu}^{T_2} \\
- & 
f_{\mu \mu}^{T_2} & f_{\mu \tau}^{T_2} \\
- & - & 0 
\end{pmatrix}M_{ee}$,\\
\vspace{0.4cm}
    $T_{3}:\begin{pmatrix}
1 & 0 & f_{e \tau}^{T_3}\\
- & f_{\mu \tau}^{T_3} & f_{\mu \tau}^{T_3}\\
- & - &  f_{\tau \tau}^{T_3} 
\end{pmatrix}M_{ee}$, \hspace{0.4cm}
$T_{4}:\begin{pmatrix}
1 & 0 & f_{e \tau}^{T_4}\\
- & f_{\mu \mu}^{T_4} & f_{\mu \tau}^{T_4} \\
- & - & f_{\mu \tau}^{T_4}
\end{pmatrix}M_{ee}$,\\
\vspace{0.4cm}
$T_{5}:\begin{pmatrix}
1 & f_{e \mu}^{T_5} & 0 \\
- & f_{\mu \mu}^{T_5} & f_{\mu \mu}^{T_5} \\
- & - &  f_{\tau\tau}^{T_5} 
\end{pmatrix}M_{ee}$, \hspace{0.4cm}
$T_{6}:\begin{pmatrix}
1 & f_{e \mu}^{T_6} & 0\\
- & f_{\mu \mu}^{T_6} & f_{\mu\tau}^{T_6} \\
- & - &f_{\mu\tau}^{T_6}
\end{pmatrix}M_{ee}$.\\
\vspace{0.4cm}
\end{center}

\noindent Eqn.(\ref{meq}) provide an important link between dark matter and $0\nu\beta\beta$ decay amplitude $\left|M_{ee}\right|$. Using Eqn.(\ref{I}), we calculate $M_{\alpha\beta}$, in Eqn.(\ref{meq}), in terms of loop functions $\Lambda_{k}$ and Yukawa couplings. All the six elements of neutrino mass matrix are calculated and given in Table \ref{Tab2} for each allowed hybrid texture $T_{1....6}$.

\begin{table}[H]
    \centering
    \begin{tabular}{|c|c|}
    \hline
      Texture   &  Constraining Equations \\
      \hline
        $T_1$      &  $\begin{array} {lcl}y_{e1}^{2}\Lambda_{1}+y_{e2}^{2}\Lambda_{2}+y_{e3}^{2}\Lambda_{3}&=& M_{ee}\\
y_{e1}y_{\mu1}\Lambda_{1}+y_{e2}y_{\mu2}\Lambda_{2}+y_{e3}y_{\mu3}\Lambda_{3}&=& f_{e\mu}^{T_1} M_{ee}\\
y_{e1}y_{\tau1}\Lambda_{1}+y_{e2}y_{\tau2}\Lambda_{2}+y_{e3}y_{\tau3}\Lambda_{3}&=&f_{e\mu}^{T_1} M_{ee}\\
y_{\mu1}^{2}\Lambda_{1}+y_{\mu2}^{2}\Lambda_{2}+y_{\mu3}^{2}\Lambda_{3}&=&0\\
y_{\mu1}y_{\tau1}\Lambda_{1}+y_{\mu2}y_{\tau2}\Lambda_{2}+y_{\mu3}y_{\tau3}\Lambda_{3}&=&f_{\mu\tau}^{T_1} M_{ee}\\
y_{\tau1}^{2}\Lambda_{1}+y_{\tau2}^{2}\Lambda_{2}+y_{\tau3}^{2}\Lambda_{3}&=& f_{\tau\tau}^{T_1} M_{ee} \end{array}$  \\
         \hline
               
    $T_2$     &   $\begin{array} {lcl}y_{e1}^{2}\Lambda_{1}+y_{e2}^{2}\Lambda_{2}+y_{e3}^{2}\Lambda_{3}&=& M_{ee}\\
y_{e1}y_{\mu1}\Lambda_{1}+y_{e2}y_{\mu2}\Lambda_{2}+y_{e3}y_{\mu3}\Lambda_{3}&=&f_{e\mu}^{T_2} M_{ee}\\
y_{e1}y_{\tau1}\Lambda_{1}+y_{e2}y_{\tau2}\Lambda_{2}+y_{e3}y_{\tau3}\Lambda_{3}&=&f_{e\mu}^{T_2}M_{ee}\\
y_{\mu1}^{2}\Lambda_{1}+y_{\mu2}^{2}\Lambda_{2}+y_{\mu3}^{2}\Lambda_{3}&=&f_{\mu\mu}^{T_2} M_{ee}\\
y_{\mu1}y_{\tau1}\Lambda_{1}+y_{\mu2}y_{\tau2}\Lambda_{2}+y_{\mu3}y_{\tau3}\Lambda_{3}&=&f_{\mu\tau}^{T_2} M_{ee}\\
y_{\tau1}^{2}\Lambda_{1}+y_{\tau2}^{2}\Lambda_{2}+y_{\tau3}^{2}\Lambda_{3}&=& 0\end{array}$  \\
\hline
       $T_3$   &  $\begin{array} {lcl}y_{e1}^{2}\Lambda_{1}+y_{e2}^{2}\Lambda_{2}+y_{e3}^{2}\Lambda_{3}&=& M_{ee}\\
y_{e1}y_{\mu1}\Lambda_{1}+y_{e2}y_{\mu2}\Lambda_{2}+y_{e3}y_{\mu3}\Lambda_{3}&=& 0 \\
y_{e1}y_{\tau1}\Lambda_{1}+y_{e2}y_{\tau2}\Lambda_{2}+y_{e3}y_{\tau3}\Lambda_{3}&=&f_{e\tau}^{T_3} M_{ee}\\
y_{\mu1}^{2}\Lambda_{1}+y_{\mu2}^{2}\Lambda_{2}+y_{\mu3}^{2}\Lambda_{3}&=&f_{\mu\tau}^{T_3} M_{ee}\\
y_{\mu1}y_{\tau1}\Lambda_{1}+y_{\mu2}y_{\tau2}\Lambda_{2}+y_{\mu3}y_{\tau3}\Lambda_{3}&=&f_{\mu\tau}^{T_3} M_{ee}\\
y_{\tau1}^{2}\Lambda_{1}+y_{\tau2}^{2}\Lambda_{2}+y_{\tau3}^{2}\Lambda_{3}&=& f_{\tau\tau}^{T_3} M_{ee}  \end{array}$ \\
      \hline
    $T_4$      & $\begin{array} {lcl}y_{e1}^{2}\Lambda_{1}+y_{e2}^{2}\Lambda_{2}+y_{e3}^{2}\Lambda_{3}&=& M_{ee}\\
y_{e1}y_{\mu1}\Lambda_{1}+y_{e2}y_{\mu2}\Lambda_{2}+y_{e3}y_{\mu3}\Lambda_{3}&=& 0\\
y_{e1}y_{\tau1}\Lambda_{1}+y_{e2}y_{\tau2}\Lambda_{2}+y_{e3}y_{\tau3}\Lambda_{3}&=&f_{e\tau}^{T_4} M_{ee}\\
y_{\mu1}^{2}\Lambda_{1}+y_{\mu2}^{2}\Lambda_{2}+y_{\mu3}^{2}\Lambda_{3}&=&f_{\mu\mu}^{T_4} M_{ee}\\
y_{\mu1}y_{\tau1}\Lambda_{1}+y_{\mu2}y_{\tau2}\Lambda_{2}+y_{\mu3}y_{\tau3}\Lambda_{3}&=&f_{\mu\tau}^{T_4} M_{ee}\\
y_{\tau1}^{2}\Lambda_{1}+y_{\tau2}^{2}\Lambda_{2}+y_{\tau3}^{2}\Lambda_{3}&=& f_{\mu\tau}^{T_4} M_{ee} \end{array} $  \\
         \hline
              
    $T_5$     &   $\begin{array} {lcl}y_{e1}^{2}\Lambda_{1}+y_{e2}^{2}\Lambda_{2}+y_{e3}^{2}\Lambda_{3}&=& M_{ee}\\
 y_{e1}y_{\mu1}\Lambda_{1}+y_{e2}y_{\mu2}\Lambda_{2}+y_{e3}y_{\mu3}\Lambda_{3}&=& f_{e\mu}^{T_5} M_{ee}\\
y_{e1}y_{\tau1}\Lambda_{1}+y_{e2}y_{\tau2}\Lambda_{2}+y_{e3}y_{\tau3}\Lambda_{3}&=&0\\
y_{\mu1}^{2}\Lambda_{1}+y_{\mu2}^{2}\Lambda_{2}+y_{\mu3}^{2}\Lambda_{3}&=&f_{\mu\mu}^{T_5} M_{ee}\\
y_{\mu1}y_{\tau1}\Lambda_{1}+y_{\mu2}y_{\tau2}\Lambda_{2}+y_{\mu3}y_{\tau3}\Lambda_{3}&=&f_{\mu\mu}^{T_5} M_{ee}\\
y_{\tau1}^{2}\Lambda_{1}+y_{\tau2}^{2}\Lambda_{2}+y_{\tau3}^{2}\Lambda_{3}&=& f_{\tau\tau}^{T_5} M_{ee}  \end{array}$ \\
         \hline
               
   $T_6$      &  $\begin{array} {lcl}y_{e1}^{2}\Lambda_{1}+y_{e2}^{2}\Lambda_{2}+y_{e3}^{2}\Lambda_{3}&=& M_{ee}\\
y_{e1}y_{\mu1}\Lambda_{1}+y_{e2}y_{\mu2}\Lambda_{2}+y_{e3}y_{\mu3}\Lambda_{3}&=& f_{e\mu}^{T_6} M_{ee}\\
y_{e1}y_{\tau1}\Lambda_{1}+y_{e2}y_{\tau2}\Lambda_{2}+y_{e3}y_{\tau3}\Lambda_{3}&=&0\\
y_{\mu1}^{2}\Lambda_{1}+y_{\mu2}^{2}\Lambda_{2}+y_{\mu3}^{2}\Lambda_{3}&=&f_{\mu\mu}^{T_6} M_{ee}\\
y_{\mu1}y_{\tau1}\Lambda_{1}+y_{\mu2}y_{\tau2}\Lambda_{2}+y_{\mu3}y_{\tau3}\Lambda_{3}&=&f_{\mu\tau}^{T_6} M_{ee}\\
y_{\tau1}^{2}\Lambda_{1}+y_{\tau2}^{2}\Lambda_{2}+y_{\tau3}^{2}\Lambda_{3}&=&f_{\mu\tau}^{T_{4}} M_{ee} \end{array} $  \\
         \hline
  
    \end{tabular}
\caption{Constraining equations relating loop factors and Yukawa couplings to the effective Majorana neutrino mass $\left|M_{ee}\right|$ for all hybrid textures $T_{1....6}$.}
\label{Tab2}
\end{table}
\noindent The corresponding expressions of $f^X_{\alpha\beta}$ can be read from Eqns.(24-30). Also $f_{ee}$ coefficients for all textures are unity.  
\noindent The coefficients, $f_{\alpha \beta}^{X}$ for all textures are as follows\\

\begin{align}
\left.
 \begin{array}{lll}
f_{e\mu}^{T_{1}}&=&\frac{\sin{2\theta_{12}} \sin{2\theta_{13}}(e^{2i\delta}c_{23}^{2}+s^{2}_{23})}{-4e^{2i\delta}B_{5}(c_{23}-s_{23}) c^{2}_{23} A_{3}+\frac{e^{i\delta}}{\csc{2\theta_{12}}} (2 e^{2i\delta}c_{23}^{2} A_{3}+s_{23}(A_{1}-\frac{2 C_{3}}{\sin{2\theta_{12}}}))},\\
f_{\mu\tau}^{T_{1}}&=&\frac{\sqrt{2}\cos(\frac{\pi}{4}+\theta_{23})(\frac{-4e^{i\delta} B_{5}}{\sec{2\theta_{23}}}+\frac{2\sin{2\theta_{12}}}{\csc{2\theta_{23}}}( s_{13}^{2} +e^{2i\delta}(\frac{\cos^{2}{\theta_{13}}}{\sin{2\theta_{23}}}+1)))}{-4e^{3i\delta}B_{5}(c_{23}-s_{23}) c^{2}_{23} A_{3}+\frac{e^{i\delta}}{\csc{2\theta_{12}}}  (2 e^{2i\delta}c_{23}^{2} A_{3}+s_{23}(A_{1}-\frac{2 C_{3}}{\sin{2\theta_{12}}}))},\\
f_{\tau\tau}^{T_{1}}&=&\frac{\sin{2\theta_{12}}(  \frac{2(c_{23}-s_{23})}{\csc^{2}{\theta_{13}}\csc^{2}{\theta_{23}}}+\frac{e^{2i\delta}}{\sec{\theta_{23}}}(\frac{C_{3}}{\sin{2\theta_{12}}}+\sin{2\theta_{23}})-e^{i\delta}B_{5}(B_{1}-A_{5}))}{-4e^{3i\delta}B_{5}(c_{23}-s_{23}) c^{2}_{23} A_{3}+\frac{e^{i\delta}}{\csc{2\theta_{12}}} (2 e^{2i\delta}c_{23}^{2} A_{3}+s_{23}(A_{1}-\frac{2 C_{3}}{\sin{2\theta_{12}}}))}, 
\end{array}
\right\}
\end{align}

\begin{align}
\left.
 \begin{array}{lll}
f_{e\mu}^{T_{2}}&=&\frac{\sin{2\theta_{12}} \sin{2\theta_{13}}(c_{23}^{2}+e^{2i\delta} s^{2}_{23})}{-4e^{2i\delta}B_{5}(c_{23}-s_{23}) s^{2}_{23}+2e^{3i\delta}A_3 B_4-e^{i\delta}c_{23}\sin{2\theta_{12}(A_2-4s_{13}^2)}},\\
f_{\mu\tau}^{T_{2}}&=&\frac{2B_4(s_{23}-c_{23})c_{23}^2+e^{2i\delta}\frac{s_{23}}{\csc{2\theta_{12}}}(A_4+\sin{2\theta_{23}})+e^{i\delta}B_5(A_5-B_1)}{-4e^{3i\delta}B_{5}(c_{23}-s_{23}) s^{2}_{23}+2e^{4i\delta}A_3 B_4-e^{2i\delta}c_{23}\sin{2\theta_{12}(A_2-4s_{13}^2)}},\\
 f_{\mu\mu}^{T_{2}}&=&\frac{\sin{2\theta_{12}}(  \frac{2(c_{23}-s_{23})}{\csc^{2}{\theta_{13}}\csc^{2}{\theta_{23}}}+\frac{e^{2i\delta}}{\sec{\theta_{23}}}(\frac{C_{3}}{\sin{2\theta_{12}}}+\sin{2\theta_{23}})-e^{i\delta}B_{5}(B_{1}-A_{5}))}{-4e^{3i\delta}B_{5}(c_{23}-s_{23}) s^{2}_{23}+2e^{4i\delta}A_3 B_4-e^{2i\delta}c_{23}\sin{2\theta_{12}(A_2-4s_{13}^2)}},
\end{array}
\right\}
\end{align}
\begin{align}
\left.
 \begin{array}{lll}
f_{e\tau}^{T_{3}}&=&\frac{\sin{2\theta_{12}} \sin{2\theta_{13}} (A_{1}+e^{2i\delta}A_{2})}{2e^{i\delta} \sin{2\theta_{12}} (2e^{2i\delta} c_{23}^2 A_{3}+s_{23}(A_4-\sin{2\theta_{23}}))+2e^{2i\delta}B_5(A_5+B_1)},\\
f_{\mu\tau}^{T_{3}}&=&\frac{4s_{23} (e^{i\delta} c_{23} s_{12}+c_{12} s_{13} s_{23})(e^{i\delta}c_{12} c_{23}-s_{12} s_{13} s_{23})}{2e^{i\delta} \sin{2\theta_{12}} (2e^{2i\delta} c_{23}^2 A_{3}+s_{23}(A_4-\sin{2\theta_{23}}))+e^{3i\delta}B_5(A_5+B_1)},\\
f_{\tau \tau}^{T_{3}}&=&\frac{ \frac{(A_{6}-2A_{5})}{\csc{2\theta_{12}} \csc_{13}^{2}}+ \frac{e^{2i\delta}}{\csc{2\theta_{12}}}  (2( \cos{3\theta_{23}} +\frac{A_{3} \cos{2\theta_{13}}}{s_{13}^{2}}+e^{i\delta}B_{5}(A_{5}+2A_{6}))-A_{6})}{2e^{i\delta} \sin{2\theta_{12}} (2e^{2i\delta} c_{23}^2 A_{3}+s_{23}(A_4-\sin{2\theta_{23}}))+2e^{3i\delta}B_5(A_5+B_1)},
\end{array}
 \right\}
 \end{align}

 

\begin{align}
\left.
 \begin{array}{lll}
f_{e\tau}^{T_{4}}&=&\frac{\sin{2\theta_{12}} s_{13} ((A_{1}+2 \cos{2\theta_{23}})+e^{2i\delta}(A_{2}-2 \cos{2\theta_{23}}))}{8e^{2i\delta}B_{5}s^{2}_{23}A_{3}-e^{i\delta} \sin{2\theta_{12}}(B_{3}-2 A_{4} s_{23})+e^{3i\delta}\sin{2\theta_{12}} s_{13}^{2}(A_{5}+A_{6})},\\
f_{\mu \tau}^{T_{4}}&=&\frac{2s_{23}(e^{2i\delta} A_{4} \sin{2\theta_{12}}+ 2 \cos^{2}{\theta_{23}} B_{4}+2e^{i\delta}\sin{2\theta_{23}} B_{5})}{8e^{3i\delta}B_{5}s^{2}_{23}A_{3}-e^{2i\delta} \sin{2\theta_{12}}(B_{3}-2 A_{4} s_{23})+e^{4i\delta}\sin{2\theta_{12}} s_{13}^{2}(A_{5}+A_{6})},\\
f_{\mu \mu }^{T_{4}}&=&\frac{\frac{2}{\csc{2\theta_{12}}}(\frac{2(2 c_{23}-s_{23})}{\csc{\theta_{13}}^{2}}-\frac{e^{2i\delta}}{\sec{\theta{23}}} (A_{4}-A_{1}-\frac{2}{\csc{2\theta_{23}}})+e^{i\delta}\frac{ s_{13}}{\tan{2\theta_{12}}}(A_{5}-2A_{6}))}{8e^{3i\delta}B_{5}s^{2}_{23}A_{3}-e^{2i\delta} \sin{2\theta_{12}}(B_{3}-2 A_{4} s_{23})+e^{4i\delta}\sin{2\theta_{12}} s_{13}^{2}(A_{5}+A_{6})},
\end{array}
\right\}
\end{align}


\begin{align}
\left.
 \begin{array}{lll}
f_{e\mu}^{T_{5}}&=&\frac{\sin{2\theta_{12}} \sin{2\theta_{13}}(A_{1}+e^{2i\delta}A_{2})}{\frac{-8e^{2i\delta} c_{13}^2}{s_{13^2}} A_{3} B_5+\frac{2e^{i\delta}}{\csc{2\theta_{12}}}( C_{1}\cos{2\theta_{13}} +C_{2} c_{23} )+e^{3i\delta} B_4 (A_{5}+B_{1})},\\
f_{\mu\tau}^{T_{5}}&=&\frac{2 c_{23}(e^{2i \delta} C_3 +2 C_4 - 2 e^{i \delta} B_5 \sin{2\theta_{23}})}{\frac{-8e^{3i\delta} c_{13}^2}{s_{13^2}} A_{3} B_5+\frac{2e^{2i\delta}}{\csc{2\theta_{12}}}( C_{1}\cos{2\theta_{13}} +C_{2} c_{23} )+e^{4i\delta} B_4 (A_{5}+B_{1})},\\
f_{\tau\tau}^{T_{5}}&=&\frac{2(\sin{2\theta_{12}}(-2c_{23}^{2} s_{13}^{2} C_{1}+e^{2i\delta}D_{1}-e^{id} B_{5}(2B_{2}+B_{1})))}{\frac{-8e^{3i\delta} c_{13}^2}{s_{13^2}} A_{3} B_5+\frac{2e^{2i\delta}}{\csc{2\theta_{12}}}( C_{1}\cos{2\theta_{13}} +C_{2} c_{23} )+e^{4i\delta} B_4 (A_{5}+B_{1})},  
\end{array}
\right\}
\end{align}

 
\begin{align}
\left.
 \begin{array}{lll}
f_{e\mu}^{T_{6}}&=&\frac{\sin{2\theta_{12}} s_{13}((A_{1}+2 \cos{2\theta_{23}})+e^{2i\delta}(A_{1}+2 \sin{2\theta_{23}}))}{\frac{4e^{3i\delta} C_{4} A_{3}}{s_{13}^{2}}-2e^{i\delta}c_{23} \sin{2\theta_{12}}(A_{2}-c_{13}^{2})-2e^{2i\delta}B_{5}(A_{5}+B_{1})},\\
f_{\mu\mu}^{T_{6}}&=&\frac{16e^{i\delta}B_{5}(B_{2}- \frac{c_{23}}{\csc{2\theta_{23}}})-\frac{4e^{2i\delta}}{\csc{2\theta_{12}}}(\frac{e^{-2i\delta}D_{2}}{ \csc{\theta_{13}}^{2}}-(B_{2}-\frac{2 \cos{2\theta_{13}}A_{3}}{s_{13}^{2}} +\frac{2 }{\csc{3\theta_{23}}}))}{\frac{16 e^{4i\delta} C_{4} A_{3}}{s_{13}^{2}}+8e^{2i\delta}c_{23} \sin{2\theta_{12}}(A_{2}-c_{13}^{2})-8e^{3i\delta}B_{5}(A_{5}+B_{1})},\\
 f_{\mu\tau}^{T_{6}}&=&\frac{4 c_{23}(c_{23} s_{12} s_{13}+e^{i\delta}\frac{\sin{2\theta_{12}}}{2})(-c_{12} c_{23} s_{13}+e^{i\delta}s_{12} s_{23})}{\frac{2 e^{4i\delta}C_{4} A_{3}}{s_{13}^{2}}-e^{2i\delta}c_{23} \sin{2\theta_{12}}(A_{2}-c_{13}^{2})-3e^{3i\delta}B_{5}(A_{5}+B_{1})},
\end{array}
\right\}
\end{align}

where 
\begin{align}
\left.
 \begin{array}{lllllllllllllllllllll}
A_{1}=1-\cos{2 \theta_{23}}-\sin{2 \theta_{23}},\\
A_{2}=1+\cos{2 \theta_{23}}+\sin{2 \theta_{23}},\\
A_{3}=s_{13}^{2}(c_{23}+s_{23}),\\
A_{4}=\cos{2 \theta_{13}}+\cos{2 \theta_{23}},\\
A_{5}=c_{23}-\cos{3 \theta_{23}},\\
A_{6}=-s_{23}+\sin{3 \theta_{23}},\\
B_{1}=s_{23}+\sin{3 \theta_{23}},\\
B_{2}=c_{23}+\cos{3 \theta_{23}},\\
B_{3}=(-1+4 \cos{2 \theta_{13}})\cos{2 \theta_{23}}+\cos{3 \theta_{23}},\\
B_{4}=\sin{2 \theta_{12}} s_{13}^{2},\\
B_{5}=\cos{2 \theta_{12}} s_{13}, \\
C_{1}=c_{23}-2 s_{23},\\
C_{2}=-\cos{2 \theta_{23}}+\sin{2 \theta_{23}},\\
C_{3}=\sin{2 \theta_{12}}(\cos{2 \theta_{13}}-\cos{2 \theta_{23}}),\\
C_{4}=\sin{2 \theta_{12}} s_{13}^{2} s^{2}_{23},\\
D_{1}=s_{23}(2(\cos{2 \theta_{23}}+s_{13}^{2})+\sin{2 \theta_{23}}),\\
D_{2}=\cos{3 \theta_{23}}+c_{23}(-1+4\sin{2 \theta_{23}}),\\
\end{array}
\right\}
\end{align}
  \begin{table}[t]
      \centering
\begin{tabular}{ |c|c|c|c| } 
\hline
Mixing angles & bfp $\pm$ 1$\sigma$ & 3$\sigma$ range \\
\hline

$\theta_{12}/^{o}$ & $33.82^{+0.78}_{-0.76}$ & $31.61\rightarrow36.27$ \\ 
\hline
$\theta_{23}/^{o}$ & $49.6^{+1.0}_{1.2}$ & $40.3\rightarrow52.4$ \\ 
\hline

$\theta_{13}/^{o}$ & $8.61^{+0.13}_{-0.13}$ & $8.22\rightarrow8.99$ \\ 
\hline

$\delta/^{o}$ & $215^{+40}_{-29}$ & $125\rightarrow392$ \\ 
\hline
\end{tabular}
\caption{Global fit data of neutrino mixing angles and CP phase $\delta$ \cite{Esteban:2018azc}.}
\label{tab33}
 \end{table}
\noindent We calculate the coefficients $f_{\alpha \beta}^{X}$ by randomly generating(with normal distribution) the mixing angles and CP phase within their allowed range using the data given in Table \ref{tab33}\cite{Esteban:2018azc}.
\section{Numerical analysis and discussion}
  In the last section, we derived analytical expressions required to find the relic density of DM($\Omega h^2$) and to relate it with neutrinoless double beta decay in a constrained scenario of hybrid textures. Once we randomly generate quartic coupling $\lambda$, lightest right-handed neutrino mass $M_1$ assuming hierarchy $M_1\leq M_2 < M_3$ with in their specified ranges given in Table \ref{tab3} and assuming $m_0\gtrsim M_1$, we calculate the loop functions $\Lambda_k$ using Eqn.(5), for each texture. Substituting (i) $\Lambda_k$'s and randomly varying the diagonal Yukawa couplings ($y_{e1},y_{\mu 2},y_{\tau 3}$) on the left-hand side (ii) randomly varying $\left|M_{ee}\right|$ in the range ($0-0.2$) eV on the right-hand side  of constraining equations given in Table \ref{Tab2}, we calculate the off-diagonal Yukawa couplings ($y_{e2},y_{e3},y_{\mu 1},y_{\mu3},y_{\tau 1},y_{\tau 2}$) restricting them in the range $0$ to  $1.2$. In addition, the LFV bound for the process $\mu\rightarrow e\gamma$ i.e., $Br(\mu\rightarrow e\gamma) \leq 4.2\times 10^{-13}$ \cite{MEG:2016leq} is, also, employed in the numerical analysis.
  
  \noindent Using Eqns.(10) and (14) along with calculated Yukawa couplings we evaluate  $a_{eff}$ and $b_{eff}$ which are further used to obtain thermal average cross section using Eqn.(15). Finally, we calculate the relic density of DM using Eqn.(18). 
  
  \begin{table}[t]
      \centering
      \begin{tabular}{|c|c|}
      \hline
       Parameter    & Range \\ \hline
        $\lambda$   & ($3-4$)$\times 10^{-9}$\\
        \hline
        $y_{e1},y_{\mu2},y_{\tau3}$ &  $0-1.2$
        \\
        \hline
        $M_1$ &100 GeV-8 TeV\\
        \hline
      \end{tabular}
      \caption{Ranges of parameters used in the numerical analysis.}
      \label{tab3}
  \end{table}
  
    \begin{table}[t]
  \centering
\begin{tabular}{|c|c|c| }
 \hline
 Texture & $M_{1}$ (TeV) & $\left|M_{ee}\right|$ (eV)\\
 \hline
$T_{1}$ & $\leq 2.27$ &$\geq 0.018$ \\ 
\hline
$T_{2}$ &  \multicolumn{2}{|l|}{disallowed by the observed value of relic density of DM($\Omega h^2$)} \tabularnewline  
 \hline
$T_{3}$ & $\leq 5.31$& $\geq 0.030$ \\
 \hline
$T_{4}$  & $\leq 3.93$& $\geq 0.015 $   \\
 \hline
$T_{5}$ & $\leq  2.77 $&  [0-0.200]   \\
 \hline
$T_{6}$ & $\leq 4.10$& $\geq0.024$ \\
 \hline
 \end{tabular}
 \caption{Bounds on DM mass $M_1$ and  $|M_{ee}|$ for allowed hybrid textures.}
 \label{T3}
 \end{table}
   
  \begin{figure}[t]
   \includegraphics[scale=0.45]{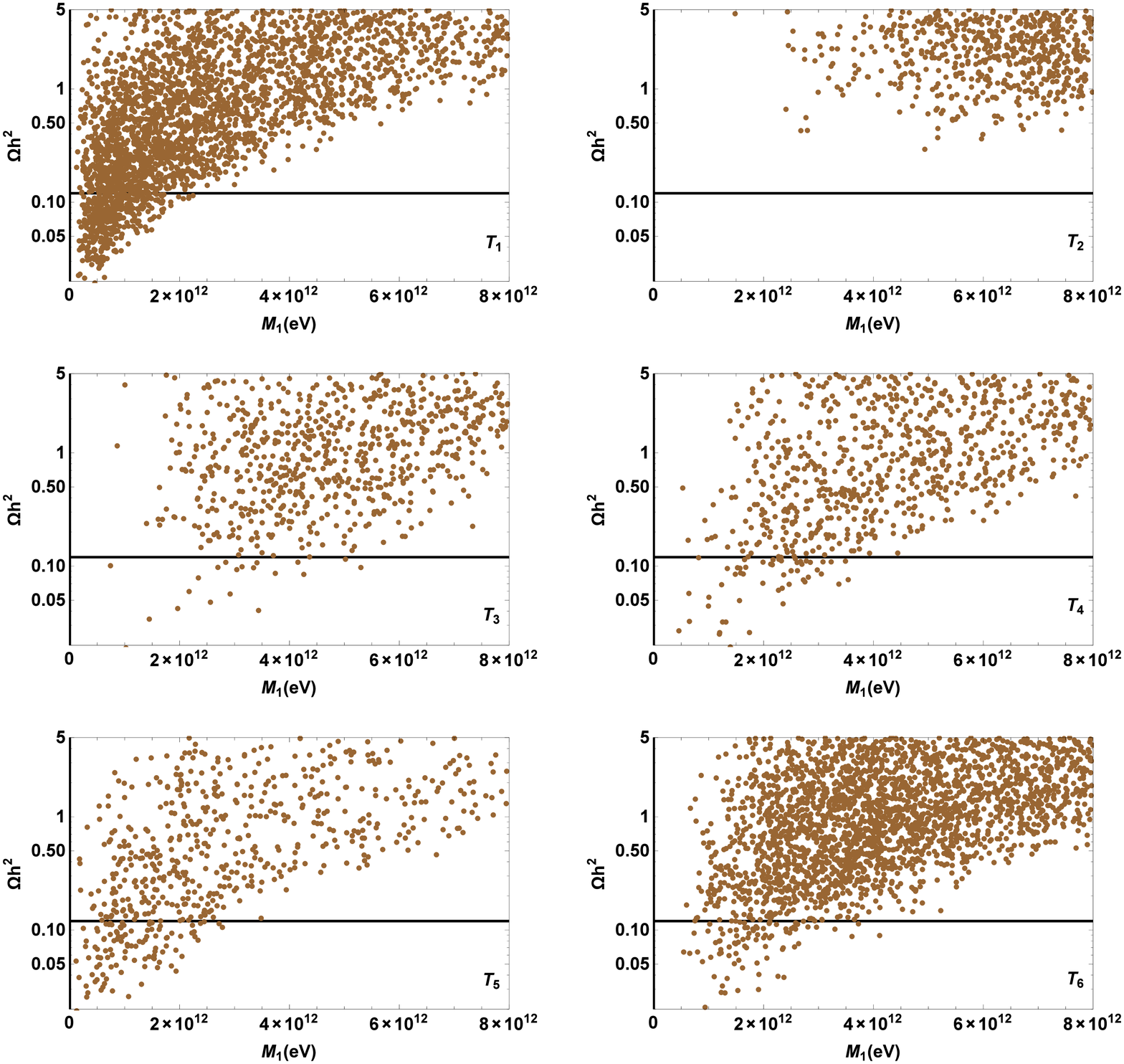}
\caption{Correlation between Relic density of DM($\Omega h^2$) and DM mass $M_1$. The horizontal line is the observed value of relic density of DM $\Omega h^2=0.1199\pm 0.0022$ \cite{Planck}.}
\label{Fig1}
\end{figure}
  
\begin{figure}[t]
   \includegraphics[scale=0.45]{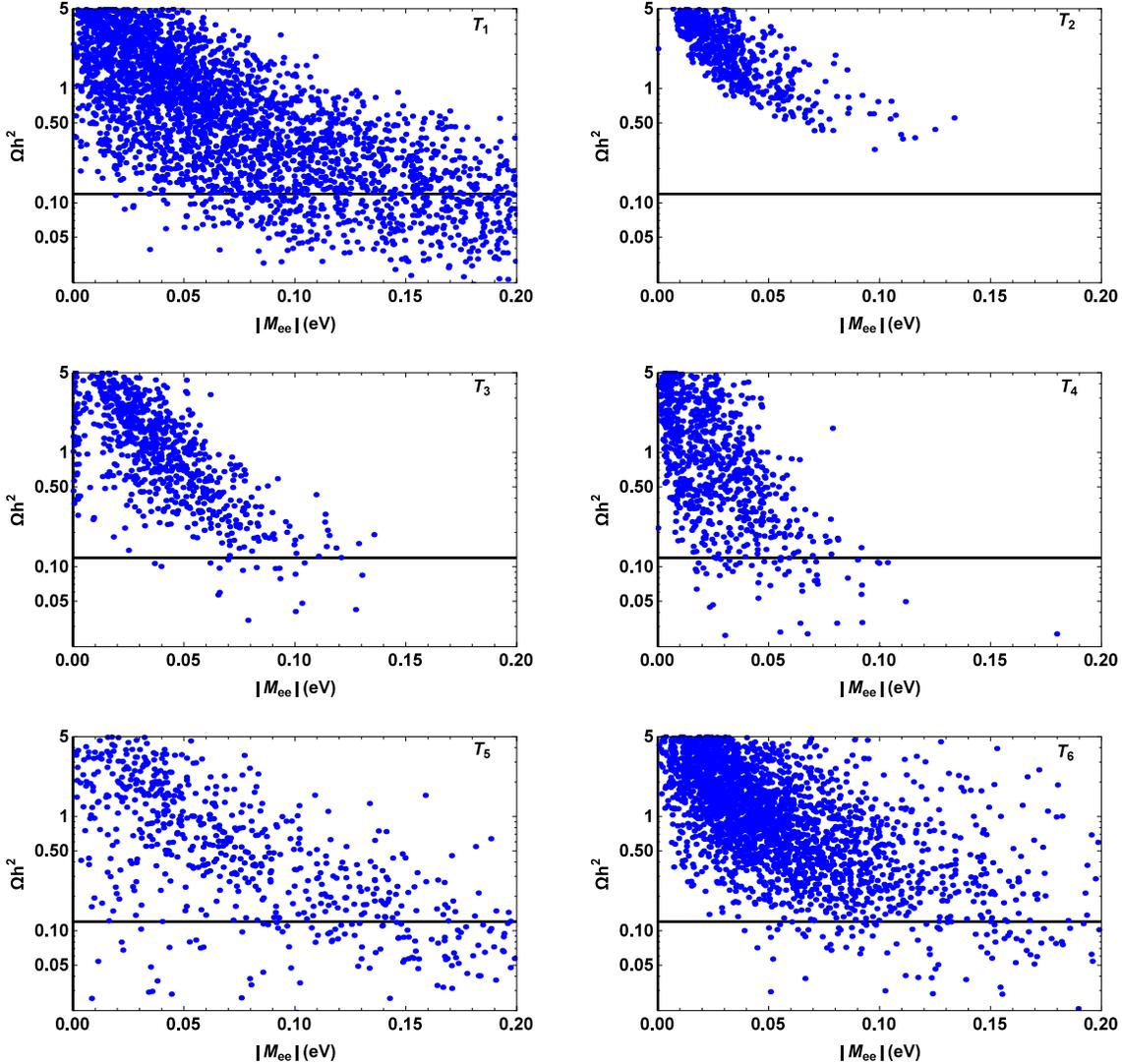}
\caption{Correlation between Relic density of DM with effective Majorana mass $\left|M_{ee}\right|$. The horizontal line is the observed value of relic density of DM $\Omega h^2=0.1199\pm 0.0022$ \cite{Planck}.}
\label{Fig2}
\end{figure}

 \noindent The correlation plots of relic density of DM($\Omega h^2$) with DM mass $M_1$ and $|M_{ee}|$ are shown in Fig.(1) and (2), respectively. The horizontal line is the observed value of relic density of DM $\Omega h^2=0.1199\pm 0.0022$\cite{Planck} which is used to constrain the DM mass $M_1$ (in Fig.(1)) and $|M_{ee}|$ (in Fig.(2)). The plots in Fig.(\ref{Fig1})  show that there is a specific range of DM mass $M_1$ which is consistent with observed value of relic density of DM. It  gives upper bound on the DM mass $M_1$ for each possible hybrid texture(Table \ref{T3}). Also, it is interesting to note that out of all these six hybrid textures, texture $T_2$ is not consistent with the observed value of relic density of DM. Therefore, we have only five possible textures which gives constrains on the DM mass $M_1$. The plots in Fig.(\ref{Fig2})  shows  the variation of relic density of DM   with $|M_{ee}|$. It is evident from Fig.(\ref{Fig2}) that there exist a lower bound on $|M_{ee}|$ for textures $T_1$, $T_3$, $T_4$ and $T_6$ while for texture $T_5$, there is no sharp lower bound and it can approach to zero.\\
  \noindent The predicted upper bounds on DM mass $M_1$ and lower bound on $|M_{ee}|$, for each hybrid texture, are given in Table \ref{T3}. The parameter space of $|M_{ee}|$ consistent with observed relic density of DM($\Omega h^2$), have imperative implications for $0\nu\beta\beta$ decay experiments. The observation and non-observations of $|M_{ee}|$
in the current and future $0\nu\beta\beta$ decay experiments\cite{Barabash:2011row,KamLAND-Zen:2016pfg,NEXT:2013wsz,NEXT:2009vsd,Licciardi:2017oqg} can further refute or validate the textures. For example, the non-observation of $0\nu\beta\beta$ decay down to the sensitivity $\mathcal{O}(0.03eV)$ will refute $T_3$ hybrid texture.

\section{Conclusions}

In this work, we have studied the implications of scotogenic model to the effective Majorana mass $|M_{ee}|$ and dark matter within the framework of hybrid textures of neutrino mass matrix. Within the formalism, we construct nine possibilities of hybrid textures having non-zero $|M_{ee}|$. Out of these nine hybrid textures, only six($T_1, T_2, T_3, T_4, T_5$ and $T_6$) are compatible with the low energy neutrino phenomenology. We derive analytic expressions to show the possible connection between non-zero elements of hybrid textures of neutrino mass matrix and $|M_{ee}|$ and employed scotogenic model to incorporate DM in our study. Using the neutrino oscillation data we calculate relic density of DM, DM mass $M_{1}$ and the effective Majorana mass $|M_{ee}|$ for allowed hybrid textures. The predicted parameter space for these quantities are shown in Fig.(\ref{Fig1}) and Fig.(\ref{Fig2}). Further, the texture $T_2$ is disallowed due to the constrain from observed relic density of DM($\Omega h^2$). The model has important implications for DM mass and $M_{ee}$. Interestingly, the DM mass $M_1$ consistent with observed relic density of DM is found to be $\mathcal{O}(1TeV)$ and is accessible at collider experiments. Specifically, for all five hybrid textures, the range of upper bound on DM mass $M_1$ is found to be ($2.27$-$5.31$)TeV.
 Fig.(\ref{Fig2})  shows the correlation between the relic density of DM($\Omega h^2$) and $|M_{ee}|$ for each hybrid texture. It is evident from Fig.(\ref{Fig2}) that except for texture $T_5$, there exist a robust lower bound on the $M_{ee}$(Table \ref{T3}) which is well within the sensitivity reach of current and future $0\nu\beta\beta$ decay experiments.

\noindent \textbf{\Large{Acknowledgments}}

 \noindent Ankush acknowledges the financial support provided by the University Grants Commission, Government of India vide registration number 201819-NFO-2018-19-OBC-HIM-75542 . M. K. acknowledges the financial support provided by Department of Science and Technology, Government of India vide Grant No. DST/INSPIRE Fellowship/2018/IF180327.


\begin{thebibliography}{99}

\bibitem{Kam11}
K. Eguchi \textit{et al.},
Phys. Rev. Lett. \textbf{90}, 021802 (2003).

\bibitem{sno1} Q. R. Ahmad \textit{et al.}, Phys. Rev. Lett. \textbf{92}, 181301 (2004).

\bibitem{Minkowski:1977sc}
P.~Minkowski,
Phys. Lett. B \textbf{67}, 421 (1977).

\bibitem{Mohapatra:1979ia}
R.~N.~Mohapatra and G.~Senjanovic,
Phys. Rev. Lett. \textbf{44}, 912 (1980).

\bibitem{TY}
T. Yanagida, Conf. Proc. C \textbf{7902131}, 95 (1979).

\bibitem{Glashow:1979nm}
S.~L.~Glashow,
NATO Sci. Ser. B \textbf{61}, 687 (1980).

\bibitem{Gell11}
] M. Gell-Mann, P. Ramond, and R. Slansky,  Conf. Proc. C \textbf{790927}, 315 (1979).


\bibitem{Ma1}
E. Ma, Phys. Rev. D \textbf{73}, 077301 (2006).

\bibitem{Ma2}
E. Ma, Phys. Rev. Lett. \textbf{81}, 1171 (1998).

\bibitem{Zee:1980ai}
A.~Zee,
Phys. Lett. B \textbf{93}, 389 (1980).

\bibitem{Ma:2001mr}
E.~Ma and M.~Raidal,
Phys. Rev. Lett. \textbf{87}, 011802 (2001).

\bibitem{Kubo:2006yx}
J.~Kubo, E.~Ma and D.~Suematsu,
Phys. Lett. B \textbf{642}, 18 (2006).

\bibitem{Hambye:2006zn}
T.~Hambye, K.~Kannike, E.~Ma and M.~Raidal,
Phys. Rev. D \textbf{75}, 095003 (2007).

\bibitem{Farzan:2009ji}
Y.~Farzan,
Phys. Rev. D \textbf{80}, 073009 (2009).


\bibitem{Ma:2007yx}
E.~Ma and U.~Sarkar,
Phys. Lett. B \textbf{653}, 288 (2007).
\bibitem{Ma:2007gq}
E.~Ma,
Phys. Lett. B \textbf{662}, 49 (2008).
\bibitem{Kajiyama:2013zla}
Y.~Kajiyama, H.~Okada and K.~Yagyu,
Nucl. Phys. B \textbf{874}, 198 (2013).
\bibitem{Aoki:2013gzs}
M.~Aoki, J.~Kubo and H.~Takano,
Phys. Rev. D \textbf{87}, 116001 (2013).
\bibitem{Kajiyama:2013rla}
Y.~Kajiyama, H.~Okada and T.~Toma,
Phys. Rev. D \textbf{88}, 015029 (2013).



\bibitem{Krauss:2002px}
L.~M.~Krauss, S.~Nasri and M.~Trodden,
Phys. Rev. D \textbf{67}, 085002 (2003).
\bibitem{Aoki:2008av}
M.~Aoki, S.~Kanemura and O.~Seto,
Phys. Rev. Lett. \textbf{102}, 051805 (2009).
\bibitem{Gustafsson:2012vj}
M.~Gustafsson, J.~M.~No and M.~A.~Rivera,
Phys. Rev. Lett. \textbf{110}, 211802 (2013).
\bibitem{Ahriche:2014cda}
A.~Ahriche, C.~S.~Chen, K.~L.~McDonald and S.~Nasri,
Phys. Rev. D \textbf{90}, 015024 (2014).
\bibitem{Ahriche:2014oda}
A.~Ahriche, K.~L.~McDonald and S.~Nasri,
JHEP \textbf{10}, 167 (2014).
\bibitem{Nomura:2016seu}
T.~Nomura and H.~Okada,
Phys. Lett. B \textbf{770}, 307 (2017).
\bibitem{Kitabayashi:2018bye}
T.~Kitabayashi,
Phys. Rev. D \textbf{98}, 083011 (2018).
\bibitem{Kitabayashi:2017sjz}
T.~Kitabayashi, S.~Ohkawa and M.~Yasu\`e,
Int. J. Mod. Phys. A \textbf{32}, 1750186 (2017).
\bibitem{Griest}
K. Griest and D. Seckel, Phys. Rev. D \textbf{43}, 3191 (1991).
\bibitem{Suematsu:2009ww}
D.~Suematsu, T.~Toma and T.~Yoshida,
Phys. Rev. D \textbf{79}, 093004 (2009).
\bibitem{Kolb:1990vq}
E.~W.~Kolb and M.~S.~Turner,
Front. Phys. \textbf{69}, 1-547 (1990).
\bibitem{Kitabayashi:2015jdj}
T.~Kitabayashi and M.~Yasu\`e,
Phys. Rev. D \textbf{93}, 053012 (2016).
\bibitem{Kitabayashi:2015tka}
T.~Kitabayashi and M.~Yasu\`e,
Int. J. Mod. Phys. A \textbf{31}, 1650043 (2016).
\bibitem{Kaneko}
S. Kaneko, H. Sawanaka and M. Tanimato, JHEP \textbf{0508}, 073 (2005).
\bibitem{Dev:2009he}
S.~Dev, S.~Verma and S.~Gupta,
Phys. Lett. B \textbf{687}, 53 (2010).
\bibitem{Goswami:2008uv}
S.~Goswami, S.~Khan and A.~Watanabe,
Phys. Lett. B \textbf{693}, 249 (2010).
\bibitem{Liu:2013oxa}
J.~Y.~Liu and S.~Zhou,
Phys. Rev. D \textbf{87}, 093010 (2013).
\bibitem{Kalita:2015tda}
R.~Kalita and D.~Borah,
Int. J. Mod. Phys. A \textbf{31}, 1650008 (2016).
\bibitem{Esteban:2018azc}
I.~Esteban, M.~C.~Gonzalez-Garcia, A.~Hernandez-Cabezudo, M.~Maltoni and T.~Schwetz,
JHEP \textbf{01}, 106 (2019).
\bibitem{Ibarra:2016dlb}
A.~Ibarra, C.~E.~Yaguna and O.~Zapata,
Phys. Rev. D \textbf{93}, 035012 (2016).
\bibitem{MEG:2016leq}
A.~M.~Baldini \textit{et al.},
Eur. Phys. J. C \textbf{76}, 434 (2016).
\bibitem{Planck}
P.~A.~R.~Ade \textit{et al.},
Astron. Astrophys. \textbf{594}, A13 (2016).

\bibitem{Barabash:2011row}
A.~S.~Barabash,
J. Phys. Conf. Ser. \textbf{375}, 042012 (2012).

\bibitem{KamLAND-Zen:2016pfg}
A.~Gando \textit{et al.},
Phys. Rev. Lett. \textbf{117}, no.8, 082503 (2016).

\bibitem{NEXT:2013wsz}
J.~J.~Gomez-Cadenas \textit{et al.},
Adv. High Energy Phys. \textbf{2014}, 907067 (2014).

\bibitem{NEXT:2009vsd}
F.~Granena \textit{et al.}, arXiv:0907.4054[hep-ex].

\bibitem{Licciardi:2017oqg}
C.~Licciardi,
J. Phys. Conf. Ser. \textbf{888}, no.1, 012237 (2017).


\end{thebibliography}
\end{document}